\documentclass[a4paper,12pt]{article}
\pdfoutput=1
\usepackage{epsfig}
\usepackage{epstopdf}
\usepackage{graphicx}
\usepackage{caption}
\usepackage{cite}
\usepackage[export]{adjustbox}
\usepackage{color}
\usepackage[dvipsnames]{xcolor}
\usepackage{slashed}
\makeatletter
\def\fmslash{\@ifnextchar[{\fmsl@sh}{\fmsl@sh[0mu]}}
\def\fmsl@sh[#1]#2{%
  \mathchoice
    {\@fmsl@sh\displaystyle{#1}{#2}}%
    {\@fmsl@sh\textstyle{#1}{#2}}%
    {\@fmsl@sh\scriptstyle{#1}{#2}}%
    {\@fmsl@sh\scriptscriptstyle{#1}{#2}}}
\def\@fmsl@sh#1#2#3{\m@th\ooalign{$\hfil#1\mkern#2/\hfil$\crcr$#1#3$}}
\makeatother
\usepackage[caption=false]{subfig}

\usepackage[sumlimits,intlimits,namelimits]{amsmath} 

\usepackage[english]{babel}

\usepackage{graphicx}

\usepackage{geometry}
\usepackage{comment}
\numberwithin{equation}{section}
\newcommand{\ecut}{E_\text{cut}}
\newcommand{\qcut}{ q^2_\text{cut}}
\newcommand{\qcutineq}{{ q^2 \ge  q^2_\text{cut}}}
\newcommand{\qq}{{ q^2}}

\begin{document}
\begin{titlepage}
\begin{flushright}
SI-HEP-2018-35 \\[0.2cm]
QFET-2018-22 \\[0.2cm]
\today
\end{flushright}

\vspace{1.2cm}
\begin{center}
{\Large\bf 
\boldmath $V_{cb}$ Determination from Inclusive $b \to c$  \unboldmath Decays: \\[2mm] 
An Alternative Method}
\end{center}

\vspace{0.5cm}
\begin{center}
{\sc Matteo Fael }, {\sc Thomas Mannel } and {\sc K. Keri Vos } \\[0.1cm]
{\sf Theoretische Physik 1, Naturwiss. techn. Fakult\"at, \\
Universit\"at Siegen, D-57068 Siegen, Germany}
\end{center}

\vspace{0.8cm}
\begin{abstract}
\vspace{0.2cm}\noindent
The determination of $V_{cb}$ relies on the Heavy-Quark Expansion and the extraction of the non-perturbative matrix elements from inclusive $b\to c$ decays. The proliferation of these matrix elements complicates 
their extraction at $1/m_b^4$ and higher, thereby limiting the $V_{cb}$ extraction. Reparametrization invariance links 
different operators in the Heavy-Quark expansion thus reducing the number of independent operators at $1/m_b^4$ to eight for the total 
rate. We show that this reduction also holds for spectral moments as long as they are defined by reparametrization invariant weight-functions. This
is valid in particular for the leptonic invariant mass spectrum ($q^2$), i.e.\ the differential rate and its moments. 
Currently, $V_{cb}$ is determined by fitting the energy and hadronic mass moments, which do not manifest this parameter reduction and depend on the full set of 13 matrix elements up to $1/m_b^4$. In light of this, we propose an experimental analysis of the $q^2$ moments to open the possibility of a model-independent $V_{cb}$ extraction from semileptonic decays including the $1/m_b^4$ terms in a fully data-driven way.
 
\end{abstract}

\end{titlepage}

\newpage
\pagenumbering{arabic}
\section{Introduction}
The Heavy Quark Expansion (HQE) has become a standard tool in the theoretical description of inclusive decays 
of heavy hadrons, allowing the derivation of precise predictions including reliable estimates of the uncertainties, see 
e.g. \cite{Manohar:2000dt}. The HQE 
for a bottom-hadron decay expresses observables as a combined series in $\alpha_s (m_b)$ and $\Lambda_{\rm QCD} / m_b$, 
where the hadronic inputs are forward matrix elements of local operators. 

One of the master application of the HQE is the determination of $V_{cb}$ from inclusive semileptonic $b \to c$ transitions, see e.g.~\cite{DinMan16} for a recent review. The extraction of $V_{cb}$ relies on the precise calculation of the total rate as well as of spectral moments, i.e.\ moments of the charged lepton energy, the hadronic mass and the hadronic energy spectra.  
Current predictions of these observables within the HQE  in $(\alpha_s (m_b))^l (\Lambda_{\rm QCD} / m_b)^k$ involve terms of order 
$(k=0,l=0,1,2)$ \cite{Jezabek:1988iv,Aquila:2005hq,Pak:2008cp,Melnikov:2008qs}, $(k=2,l=0,1)$ \cite{Becher:2007tk,Alberti:2013kxa,Mannel:2014xza} and $(k=3,4,5,l=0)$, while $k=1$ vanishes for all $l$. Using the experimental data on the total rate and on the energy and hadronic mass moments \cite{Acosta:2005qh,Csorna:2004kp,Abdallah:2005cx,Aubert:2004td,Aubert:2009qda, Urquijo:2006wd,Schwanda:2006nf}
to obtain the hadronic parameters allows the extraction of $| V_{cb}|$ with a relative precision of about 2\% \cite{Gam13, Alb14}. This error includes an additional 1.4\% theoretical uncertainty due  to the missing higher-order corrections in the expression for the width~\cite{Benson:2003kp,Gambino:2011cq,Gam13}.

With this current strategy, a  model independent, meaning a fully data-driven determination of $V_{cb}$, including the extraction of the HQE parameters from the data,  is possible only up to $1/m_b^3$; up to this order there are only four independent hadronic parameters. 
Starting at order $1/m_b^4$, their proliferation  complicates the extraction 
from data. Therefore, one has to resort to modelling the higher-order terms in order to at least get a quantitative picture of their possible size. Such a model approach was suggested in \cite{Man10, Heinonen:2014dxa}, where the Lowest State Saturation Ansatz (LSSA) was used to estimate the effect of the orders $1/m_b^4$ and higher. This study indicates that such higher-order terms are likely negligible at the current level of precision. This was confirmed by a global fit\cite{Gam16}, where the LSSA was used to provide loose constraints on the higher-order matrix elements. A sub-percent reduction in $V_{cb}$ was found. However, these statements depend to a large extend on the LSSA, which is an ad hoc ansatz, and thus it is desirable to validate the smallness of  the $1/m_b^4$ terms by a model-independent approach.

In this paper we propose an alternative method for a $V_{cb}$ extraction which still makes use of the HQE, but in a slightly different set-up. 
It is known that reparametrization invariance (RPI), a symmetry within the HQE reflecting Lorentz invariance of the underlying 
QCD \cite{Bra03,Heinonen:2012km,Berwein:2018fos}, induces relations between the coefficients of HQE parameters
  \cite{Chen:1993np,Luke:1992cs,Manohar10,Gun17,Kobach:2018pie}. Recently, two of us discussed that these relations lead to a reduction of independent parameters for specific observables, in particular for the total rates \cite{MannelVos}. Phrased differently, these observables depend only on specific linear combinations of the most general set of HQE parameters. This reduced set of parameters involves only three elements up to $1/m_b^3$ (including the chromomagnetic parameter, which can be extracted from spectroscopy as well) and only five additional inputs once $1/m_b^4$  are terms are included.  

For the alternative $V_{cb}$ determination, the observable we propose is the leptonic invariant mass $(q^2)$ spectrum and, 
more specifically,  the moments of this spectrum. In the next section we give a short reprise of the findings in \cite{MannelVos} and we examine 
the consequences for observables other than total rates in section 3. In section 4, we use this reasoning to compute the $q^2$ spectrum and its moments. Finally, in section 5 we discuss the alternative extraction, in particular the possibility to push the $V_{cb}$ extraction to order $1/m_b^4$ 
without making use of models for the HQE parameters, i.e.\ to have a fully data-driven analysis up to this order.

\section{Reparametrization Invariance of the HQE}
The HQE for semileptonic $b \to c $ decays is set up starting from the time-ordered product of two weak currents 
\begin{equation} \label{eq:intro}
R_{\mu\nu}(q) = \int d^4 x \, e^{iq \cdot x} \, T[ \bar{b}(x)  \Gamma_\nu c(x)  \, \bar{c} (0) \bar\Gamma_\mu  b(0) ]  \, , 
\end{equation} 
where $\Gamma_\mu = \gamma_\mu (1-\gamma_5)$. In order to set up an $1/m_b$ expansion we re-define the $b$ quark field operators as 
\begin{equation} \label{phaseredef} 
b(x) = \exp(-i m_b (v \cdot x)) b_v (x)   \ ,
\end{equation} 
which results in 
\begin{equation} \label{eq:intro2}
R (S) = \int d^4 x \, e^{-i m_b (S \cdot x)} \, T[ \bar{b}_v(x)  \Gamma c(x)  \, \bar{c} (0) \bar\Gamma  b_v (0) ]  \, , 
\end{equation}
where here and in the following we have suppressed the indices for simplicity, and we define $S = v-q/m_b$.  

The next step is to perform an Operator Product Expansion (OPE) for the time-ordered product:
\begin{equation}  \label{OPE}  
R(S)  =  \sum_{n=0}^\infty  C_{\mu_1 \cdots \mu_n}^{(n)} (S)  \otimes \bar{b}_v (iD^{\mu_1} \ldots iD^{\mu_n})  b_v \, ,
\end{equation}  
where the symbol $\otimes$ is a shorthand notation for the proper contraction of the spinor indices of the coefficient $C$ with the 
ones of the quark fields. These $C$ coefficients depend on $1/m_b^{n+3}$, assuming $R(S)$ to be dimensionless. Taking the forward matrix element 
\begin{equation}
\langle \bar{b}_v \ldots b_v \rangle \equiv \langle B(p) |\bar{b}_v \ldots b_v | B(p) \rangle  \ ,
\end{equation} 
we obtain the hadronic correlator
\begin{equation}
T = \langle  R (S)  \rangle \ .
\end{equation}
Via the optical theorem, $T $ yields the hadronic tensor for the inclusive 
transition $B \to X_c \ell \bar{\nu}$ :
\begin{equation} \label{eq:Wope}
   \, W (p,q) = -\frac{1}{\pi}{\rm Im} \, \langle  R(S)  \rangle  = {\rm Im}  \sum_{n=0}^\infty  
\, C_{\mu_1 \ldots \mu_n}^{(n)} (S)  \otimes \langle \bar{b}_v (iD^{\mu_1} \ldots iD^{\mu_n})  b_v  \rangle \ .
\end{equation} 

The key observation is that both (\ref{eq:intro2}) as well as its OPE (\ref{OPE}), are independent of $v$, as long as all orders in the OPE are taken into account. This means that both are invariant under the reparametrization (RP) transformation $\delta_{\rm RP} $ that shifts $ v_\mu \longrightarrow v_\mu + \delta v_\mu $. In fact, the transformation rules 
\begin{eqnarray}
&& \delta_{\rm RP} \, v_\mu = \delta v_\mu  \quad  \mbox{with} \quad v \cdot \delta v = 0,   \label{RPT1}\\
&& \delta_{\rm RP} \, i D_\mu = - m_b \delta v_\mu , \label{RPT2}\\ 
&& \delta_{\rm RP} \, b_v (x) =   i m_b (x \cdot \delta v) b_v (x) ,  \quad \mbox{in particular} \quad \delta_{\rm RP} \,  b_v (0) = 0 \ ,
 \label{RPT3}
\end{eqnarray}   
show that reparametrization invariance (RPI), which dictates also that $\delta_{\rm RP}R(S)=0$, connects subsequent orders in the $1/m_b$ series of Eq.~(\ref{OPE}). 
This generates the well known relations between the coefficients $C$ at order $n$ and $n+1$~\cite{MannelVos}: 
\begin{equation}
  \delta_\mathrm{RP} C^{(n)}_{\mu_1 \dots \mu_n}(S) =
  m_b \, \delta v^\alpha 
  \Big[ 
  C^{(n+1)}_{\alpha \mu_1 \dots \mu_n}(S)
  +C^{(n+1)}_{\mu_1 \alpha \dots \mu_n}(S)
  + \dots +
  C^{(n+1)}_{\mu_1 \dots \mu_n \alpha}(S)
  \Big] \, .
  \label{eqn:RPIrelation}
\end{equation}
In turn, the hadronic matrix elements $\langle \bar{b}_v (iD_{\mu_1} \cdots iD_{\mu_n})  b_v  \rangle $ can be expressed in terms of scalar matrix elements, such as the kinetic energy parameter $\mu_\pi^2$ and the chromomagnetic parameter $\mu_G^2$ at $n = 2$. However the number of independent parameters grows factorially in the $1/m_b$ expansion (at tree level there are nine and 18 at order $1/m_b^4$ and $1/m_b^5$, respectively~\cite{Man10,Das06}) and therefore their extraction from data becomes challenging already at order $1/m_b^4$. 

Due to RPI, as discussed in \cite{MannelVos}, the total rate depends only on a restricted set of parameters, which are given by fixed linear combination of the matrix elements defined for the general case. To this end, up to order $1/m_b^4$ there are only eight independent parameters at tree level, defined by~\cite{MannelVos}:
\begin{subequations}\label{eq:MEs}
\begin{eqnarray}
&& \langle \bar{b}_v b_v \rangle = 2 m_B \mu_3 \vphantom{\frac{1}{1} } \ ,\\
&& \langle \bar{b}_v  (i D_\alpha) (i D_\beta) (-i \sigma^{\alpha \beta} ) b_v \rangle =  2m_B \mu_G^2  \vphantom{\frac{1}{1} } \ , \\ 
&& \frac{1}{2}\langle \bar{b}_v \left[ (iD_\mu) \, , \,   \left[ \left( i vD + \frac{1}{2m_b} (iD)^2 \right)  \, , \, (i D^\mu) \right] \right]   b_v \rangle 
= 2 m_B \tilde\rho_D^3 \ ,\\ 
&& \langle \bar{b}_v \left[ (iD_\mu) \, , \,  (iD_\nu) \right]  \left[ (iD^\mu)  \, , \, (i D^\nu) \right]   b_v \rangle 
= 2 m_B r_{G}^4  \vphantom{\frac{1}{1} }\ ,  \\ 
\label{eq:rE4}&& \langle \bar{b}_v \left[ (ivD ) \, , \,  (iD_\mu) \right]  \left[ (ivD)  \, , \, (i D^\mu) \right]   b_v \rangle 
= 2 m_B r_{E}^4  \vphantom{\frac{1}{1} } \ , \\
&& \langle \bar{b}_v \left[ (iD_\mu) \, , \,  (iD_\alpha) \right]  \left[ (iD^\mu)  \, , \, (i D_\beta) \right]   (-i \sigma^{\alpha \beta})  b_v \rangle 
= 2 m_B s_{B}^4  \vphantom{\frac{1}{1} }\ ,  \\ 
&& \langle \bar{b}_v \left[ (ivD) \, , \,  (iD_\alpha) \right]  \left[ (ivD)  \, , \, (i D_\beta) \right]   (-i \sigma^{\alpha \beta})  b_v \rangle 
= 2 m_B s_{E}^4  \vphantom{\frac{1}{1} }\ ,  \\ 
&& \langle \bar{b}_v  \left[ iD_\mu \, , \, \left[ iD^\mu \, , \,  \left[ iD_\alpha \, , \, iD_\beta \right] \right] \right]  (-i \sigma^{\alpha \beta}) b_v 
\rangle = 2 m_B s_{qB}^4  \vphantom{\frac{1}{1} } \ ,
\label{eqn:sqB}
\end{eqnarray} 
\end{subequations}
 Here we have redefined $\rho_D^3$ to include its RPI completion as discussed in Ref.~\cite{MannelVos} (see Eq.~\eqref{eq:rhodexp}). 

In the following we discuss under which conditions observables other than 
the total rate can be expressed in terms of this reduced set of parameters. 

\section{Generalized Moments}
\label{sec:genmom}
To obtain the semileptonic decay rate we have to multiply the hadronic tensor $W(p,q)$ by the leptonic 
tensor $L(k,k')$ which depends on the charged lepton momentum $k$ and the neutrino momentum $k'$.
The observables we will consider are generalized moments, which are defined as 
\begin{equation}
\langle M[w] \rangle = \int \frac{d^4 q}{(2\pi)^4} \widetilde{dk} \widetilde{dk'}   \, w(v,k,k') \langle R(S) \rangle L(k,k')
(2 \pi)^4 \delta^4 (q-k-k')
\end{equation}  
where $\widetilde{dk}$, $\widetilde{dk'}$  denote the usual phase space elements and 
$w(v,q)$ is a (smooth) weight function of the kinematic variables $k,k'$ and the vector $v$. As an example, 
$w = (v \cdot k)^n$ yields the (unnormalized) $n^{\rm th}$ charged lepton energy moment, once the velocity 
vector $v$ is identified with the velocity of the decaying $B$ meson.

In analogy to $R(S)$, we assume that the operator $M$ has an OPE according to 
\begin{equation}  \label{opeM}
M[w]  =  \sum_{n=0}^\infty  a_{\mu_1 \cdots \mu_n}^{(n)}   \otimes \bar{b}_v (iD^{\mu_1} \cdots iD^{\mu_n})  b_v  \ .
\end{equation} 
Applying now the RP transformation to (\ref{opeM}), gives a similar relation as for the total rate, except that the left-hand side of (\ref{opeM}) becomes: 
\begin{equation}
\delta_{\rm RP}  M[w]  =  \int \frac{d^4 q}{(2\pi)^4} \widetilde{dk} \widetilde{dk'}  \, \left[ \delta_{\rm RP} w(v,k,k') \right]   R(s)   
 L(k,k') (2 \pi)^4 \delta^4 (q-k-k')  \ .
\end{equation} 
We observe that for reparametrization-invariant weight-functions
$$
\delta_{\rm RP} w(v,k,k') = 0 \, , 
$$ 
(which is the case if $w$ does not depend on $v$), we obtain  for the coefficients $a^{(n)}$ the exact same relation~\eqref{eqn:RPIrelation} as for the total rate. 
Therefor, for reparametrization-invariant observables we will have the same reduction of HQE parameters as for the total rates. 
For the semileptonic decays considered here, the leptonic invariant mass ($q^2$) spectrum has this property, since the corresponding weight function, 
$$
w(v,k,k') = \delta (q^2 - (k+k')^2) \, ,  
$$
is manifestly $v$ independent. 

Before discussing the $q^2$ spectrum in more detail, we consider weight functions which are not reparametrization invariant. In such cases we obtain the relation:
\begin{eqnarray}
\delta_{\rm RP}  M[w]  &=&   \int \frac{d^4 q}{(2\pi)^4} \widetilde{dk} \widetilde{dk'}  \, \left[ \delta_{\rm RP} w(v,k,k') \right]    
 L(k,k') (2 \pi)^4 \delta^4 (q-k-k')   R(s) 
\\ \nonumber 
&=& 
\sum_{n=0}^\infty   \left[  \int \frac{d^4 q}{(2\pi)^4} \widetilde{dk} \widetilde{dk'}  \, \left[ \delta_{\rm RP} w(v,k,k') \right]    
 L(k,k') (2 \pi)^4 \delta^4 (q-k-k')      C_{\mu_1 \cdots \mu_n}^{(n)} (S)  
\right] \\ \nonumber && \qquad \qquad \otimes \,  \bar{b}_v (iD^{\mu_1} \cdots iD^{\mu_n})  b_v  
\\ \nonumber 
&=& \sum_{n=0}^\infty  \left[ \delta_{\rm RP} a_{\mu_1 \cdots \mu_n}^{(n)} \right]   \otimes \bar{b}_v (iD^{\mu_1} \cdots iD^{\mu_n})  b_v  
\nonumber \\ \nonumber 
&& - m   \sum_{n=0}^\infty  a_{\mu_1 \cdots \mu_n}^{(n)}   \otimes  
 \left[ \delta v^{\mu_1} \, \bar{b}_v (i D^{\mu_2}) \cdots  (i D^{\mu_n}) b_v  
  + \delta v^{\mu_2} \, \bar{b}_v (i D^{\mu_1})(i D^{\mu_3}) \cdots  (i D^{\mu_n})  b_v   \right. \nonumber \\  
&& \left. \qquad \qquad \qquad  \cdots + \delta v^{\mu_n} \, \bar{b}_v (i D^{\mu_1}) \cdots  (i D^{\mu_{n-1}})  b_v \right]  \nonumber
\end{eqnarray} 
from which we get a relation for the $a^{(n)}$ of the form 
\begin{align} \label{RPI1}
\delta_{\rm RP} a_{\mu_1 \cdots \mu_n}^{(n)} &= 
m_b \, \delta v^{\alpha}  \left( a_{\alpha \mu_1 \cdots \mu_n}^{(n+1)}  +  a_{ \mu_1 \alpha \mu_2 \cdots \mu_n}^{(n+1)}   
+ \cdots + a_{ \mu_1 \cdots \mu_n \alpha}^{(n+1)} \right)  + g^{(n)}_{\mu_1 \cdots \mu_n}   & n &= 0,1,2, \dots
\end{align} 
with the ``inhomogeneous'' term
\begin{equation}
 g^{(n)}_{\mu_1 \cdots \mu_n}  = 
 \left[  \int \frac{d^4 q}{(2\pi)^4} \widetilde{dk} \widetilde{dk'}  \, \left[ \delta_{\rm RP} w(v,k,k') \right]    
 L(k,k') (2 \pi)^4 \delta^4 (q-k-k')      C_{\mu_1 \cdots \mu_n}^{(n)} (S)  
\right] \ .
\end{equation} 
This term eventually requires the introduction of the complete set of HQE parameters at least in the general case, as it happens for the lepton energy and hadronic invariant mass spectrum and the forward-backward asymmetry proposed in~\cite{Turczyk:2016kjf}.
It remains to be seen, if one can reduce the parameter set even for weight functions that are not reparametrization invariant.  

 \section{The \boldmath $q^2$  Spectrum and its Moments}
 \label{sec:q2moments}
The differential $q^2$ spectrum and the $q^2$ moments are most easily obtained by first integrating the triple differential decay rate over the lepton energy $E_\ell$. The double differential decay rate can then be expressed in terms of the scalar functions $W_i$ as
\begin{equation}
  \frac{d^2\Gamma}{d\hat{q}^2 ds} = 96 \, {  \Gamma_0 } \, \sqrt{s^2-\hat{q}^2} \theta(s-\sqrt{\hat{q}^2}) \theta(\hat{q}^2) 
 \left[
   \hat{q}^2 W_1(\hat{q}^2, s) 
   +\frac{1}{3}(s^2 - \hat{q}^2) W_2(\hat{q}^2, s) 
 \right] \, ,
\end{equation}
\begin{equation}
\Gamma_0 = \frac{G_F^2 m_b^5 |V_{cb}|^2}{192 \pi^3} \ .
\end{equation}
The normalized variables are defined as
\begin{equation}
 \hat{q}^2 \equiv \frac{q^2}{m_b^2}\ , \;\;
 s \equiv \frac{v\cdot   q }{m_b} \ , \;\; 
 \rho= \frac{m_c^2}{m_b^2}\ ,
\end{equation}
and the functions $W_i(\qq,s)$ are the imaginary parts of the Lorentz decomposed hadronic correlator in Eq.~\eqref{eq:Wope}:
\begin{eqnarray*}
  W_{\mu\nu} =
  -g_{\mu\nu} W_1
  +v_\mu v_\nu W_2
  -i \varepsilon_{\mu\nu\rho\sigma} v^\rho q^\sigma W_3
  +q_\mu q_\nu W_4
  +(q_\mu v_\nu+q_\nu v_\mu) W_5 \, .
\end{eqnarray*}
where the optical theorem relates $W_i = - \text{Im} T_i/\pi$. 
At tree-level, the required $T_i$ are most conveniently derived following \cite{Man10,Nov83, Das06} by introducing a background field propagator
\begin{equation}
S_{\rm BGF} = \frac{1}{m_b \slashed{S} + i\slashed{D} - m_c} \ ,
\end{equation} 
that must be evaluated including the necessary Dirac matrices for the hadronic current. The expression up to order $1/m_b^n$ is obtained by expanding this propagator according to 
\begin{equation}
S_{\rm BGF} = \frac{1}{m_b \slashed{S}-m_c} \sum_{n=0}^\infty  \left( (i\slashed{D}) \frac{-1}{m_b \slashed{S}-m_c} \right)^n 
\end{equation}
up to the desired order in the residual momentum $(i\slashed{D})$. 
Forward matrix elements containing strings of covariant derivatives must be expressed in terms of scalar matrix elements.
To this end, thanks to the equations of motion, dedicated trace formulas can be derived to project matrix elements of the form
  $
    \bar b_v (i D^{\mu_1} \dots i D^{\mu_n} \Gamma) b_v
  $
   onto the complete set of HQE operators, i.e.\ the RPI parameters in~(\ref{eq:MEs}) as well as the redundant ones in~(\ref{eq:MEsad}).
The required $T_i$ are eventually obtained in terms of the full set of operators and inverse powers of the propagator $\Delta_0$:
\begin{equation}
\Delta_0 = m_b^2 -m_c^2 +  q^2 - 2 m_b v \cdot q \ .
\end{equation}
The imaginary part can be obtained from the $T_i$ using the relation 
\begin{equation}
 -\frac{1}{\pi} \text{Im}\left(\frac{1}{\Delta_0}\right)^{n+1} = 
 \frac{(-1)^n}{n!} \frac{1}{(m_b^2)^{n+1}}\delta^{(n)}(1-\rho+\hat q^2 - 2 s) \, .
\end{equation}
Finally, the integration over $s$ gives the differential rate in terms of $\delta^{(n)} (\hat z)$ where
 \begin{equation}
\hat  z\equiv  1 - 2 \sqrt{\hat{q}^2} + \hat{q}^2 - \rho \ .
 \end{equation}
We proved that indeed, the differential $q^2$ spectrum only depends on the reduced set of matrix elements in Eq.~\eqref{eq:MEs}. The same reduction holds also for the (normalized) $\hat{q}^2$ moments defined as
\begin{equation}
\mathcal{Q}_n  \equiv
 \frac{1}{\Gamma_0} \int_{\hat{q}^2=0}^\infty d\hat{q}^2 (\hat{q}^2)^n \frac{d\Gamma}{d\hat{q}^2} \ .
 \label{eqn:Qcal}
\end{equation}
We have verified this explicitly by calculating the moments up to $n=4$. 
Their lengthy expressions are attached  to  the  arXiv  version of this paper. However, we can present them in a compact form by taking the limit $m_c \to 0$, i.e.\ keeping only the non-vanishing terms in the $\rho \to 0$ limit:
\begin{align}
  \mathcal{Q}_1 {}& = 
  \frac{3}{10} \mu_3
  -\frac{7}{5} \frac{\mu_G^2}{m_b^2}
  +\frac{\tilde \rho_D^3}{m_b^3}
  \left( 19+8 \log \rho \right)
  -\frac{r_E^4}{m_b^4}
  \left( \frac{1292}{45}+\frac{40}{3} \log \rho  \right)
  -\frac{s_B^4}{m_b^4}
  \left( 8+2\log \rho \right) \notag \\[5pt]
  &+\frac{13}{120}\frac{s_{qB}^4}{m_b^4}
  +\frac{s_E^4}{m_b^4}
  \left( \frac{63}{5}+4\log \rho \right)
  +\frac{r_G^4}{m_b^4}
  \left( \frac{827}{45}+\frac{22}{3}\log \rho \right), \\[5pt]
  \mathcal{Q}_2{}& =
  \frac{2}{15} \mu_3
  -\frac{16}{15} \frac{\mu_G^2}{m_b^2}
  +\frac{\tilde \rho_D^3}{m_b^3}
  \left( \frac{358}{15}+8 \log \rho \right)
  -\frac{r_E^4}{m_b^4}
  \left( \frac{2888}{45}+\frac{64}{3} \log \rho  \right)
  -\frac{s_B^4}{m_b^4}
  \left( \frac{259}{15}+4\log \rho \right) \notag \\[5pt]
  &+\frac{s_{qB}^4}{m_b^4}
  \left( \frac{251}{180}+\frac{1}{3} \log \rho \right)
  +\frac{s_E^4}{m_b^4}
  \left( \frac{908}{45}+\frac{16}{3}\log \rho \right)
  +\frac{r_G^4}{m_b^4}
  \left( \frac{1373}{45}+\frac{28}{3}\log \rho \right), \\[5pt]
  \mathcal{Q}_3{}& =
  \frac{1}{14} \mu_3
  -\frac{6}{7} \frac{\mu_G^2}{m_b^2}
  +\frac{\tilde \rho_D^3}{m_b^3}
  \left( \frac{2888}{105}+8 \log \rho \right)
  -\frac{r_E^4}{m_b^4}
  \left( \frac{33098}{315}+\frac{88}{3} \log \rho  \right) \notag \\[5pt]
  &-\frac{s_B^4}{m_b^4}
  \left( \frac{5867}{210}+6\log \rho \right) 
  +\frac{s_{qB}^4}{m_b^4}
  \left( \frac{3763}{1260}+\frac{2}{3} \log \rho \right) \notag \\[5pt]
  &+\frac{s_E^4}{m_b^4}
  \left( \frac{1787}{63}+\frac{20}{3}\log \rho \right)   
  +\frac{r_G^4}{m_b^4}
  \left( \frac{27373}{630}+\frac{34}{3}\log \rho \right), \\[5pt]
  \mathcal{Q}_4 & =
  \frac{3}{70} \mu_3
  -\frac{5}{7} \frac{\mu_G^2}{m_b^2}
  +\frac{\tilde \rho_D^3}{m_b^3}
  \left( \frac{213}{7}+8 \log \rho \right)
  -\frac{r_E^4}{m_b^4}
  \left( \frac{47252}{315}+\frac{112}{3} \log \rho  \right)\notag \\[5pt]
  &-\frac{s_B^4}{m_b^4}
  \left( \frac{1389}{35}+8\log \rho \right)   
  +\frac{s_{qB}^4}{m_b^4}
  \left( \frac{4031}{840}+ \log \rho \right) \notag \\[5pt]
  &+\frac{s_E^4}{m_b^4}
  \left( \frac{3893}{105}+8\log \rho \right)
  +\frac{r_G^4}{m_b^4}
  \left( \frac{17978}{315}+\frac{40}{3}\log \rho \right).
\end{align}
These expressions in the massless limit can be used for $b\to u$ transitions by replacing $\log\rho \to \log (\mu^2/m_b^2)$. In this case four-quark operators contribute as well to the OPE expansion in Eq.~\eqref{OPE}.

\section{An Alternative Method to Determine \boldmath $V_{cb}$}
\label{sec:newmethod}
As discussed, the $q^2$ moments are Lorentz invariant and RPI and therefore up to $1/m_b^4$, they only depend on the reduced set of eight parameters
given in Eq.~\eqref{eq:MEs}. Currently, $V_{cb}$ is extracted from inclusive decays by fitting the HQE matrix elements to the experimental data of the electron energy and hadronic invariant mass moments. However, these quantities are not RPI, and therefore depend on the full set of matrix elements. 

Moreover, from the experimental side momentum cuts need to be implemented, since the full phase space cannot be covered. To this end, one may either extrapolate using the theoretical expression for the differential rate, or one needs to take into account the cut in the theoretical prediction. Specifically, the analysis of the charged lepton energy and hadronic invariant mass require a cut on the charged lepton energy. The moments including such a cut are defined as 
\begin{equation} \label{cutmom}
  \left\langle x^n \right\rangle_{E_\ell > \ecut} = 
  \frac{
    \int_{E_\ell > \ecut} dx \, x^n \, \frac{d\Gamma}{dx}
  }{
     \int_{E_\ell > \ecut} dx \, \frac{d\Gamma}{dx}
  } ,
\end{equation}
with $x=E_\ell$, for the electron energy moments, and $x=M_X$ for the hadronic invariant mass moments. 

The fit performed in \cite{Gam13, Alb14,Gambino:2004qm} makes use of these moments, including the energy cut. In fact, the moments up to $n = 4$ and their computable cut-dependence allow for a fully data-driven analysis up to $1/m_b^3$, which means that $V_{cb}$, the quark masses as well as the HQE parameters $\mu_\pi^2, \mu_G^2, \rho_D^3$ and $\rho_{LS}^3$ can be fitted from data. 
Accessing higher order in the $1/m_b$ expansion requires to model the HQE parameters starting at $1/m_b^4$; this has been investigated  in \cite{Gam16} using the LSSA proposed in~\cite{Man10, Heinonen:2014dxa}. 

We therefore propose to determine $V_{cb}$ from the moments in ${q}^2$. As these moments only depend on the reduced number of independent HQE parameters, this would allow for a fully data-driven extraction of $V_{cb}$ up to $1/m_b^4$. However, the $q^2$ moments as defined in (\ref{cutmom}), i.e.\ inserting $x={q}^2$, will depend on the complete set of parameters, since the electron 
energy $v \cdot k$ is not an RPI quantity. This can be solved either by removing the cut by extrapolating the data to the 
full phase space, or by implementing a cut in an RPI quantity, such as ${q}^2$.

\subsection{Employing a \boldmath ${q^2}$ cut \unboldmath} 
In the following, we further discuss the implementation and concequences of a cut in ${q}^2$. 
We define the moments as follows:
\begin{equation}
  \left\langle (q^2)^n \right\rangle_\qcutineq \equiv
    \left. \int_{\qcut}^{m_b^2(1-\sqrt{\rho})^2} d\qq \, (\qq)^n \, \frac{d\Gamma}{d\qq}
    \right/
     \int_{\qcut}^{m_b^2(1-\sqrt{\rho})^2} d\qq \, \frac{d\Gamma}{d\qq}  =
     (m_b^2)^n\frac{\mathcal{Q}_n (\hat q^2_\mathrm{cut})}
     { \mathcal{Q}_0(\hat q^2_\mathrm{cut})} \, ,
     \label{eqn:q2momdef}
\end{equation}
where the $\mathcal{Q}_n(\hat q^2_\mathrm{cut})$ are defined as in Eq.~\eqref{eqn:Qcal}, but with $\hat q^2_\mathrm{cut}$ as lower limit of integration instead of $\hat q^2 = 0$. We also define the fraction $R^*$:
\begin{equation}
R^* \equiv \frac{\Gamma_\qcutineq}{\Gamma_{\rm tot}} = \,
\frac{\mathcal{Q}_0 (\hat q^2_\mathrm{cut})}
     { \mathcal{Q}_0} ,
     \label{eqn:Rstar}
\end{equation}
The explicit expressions of $\mathcal{Q}_n(\hat q^2_\mathrm{cut})$, with $n=0,\dots,4$ are also given as an ancillary file.
Similar as for the charged lepton energy moments and hadronic invariant mass moments, central moments are less correlated, therefore we define  
\begin{align}\label{eq:cenmom}
  q_1\equiv \left\langle  q^2 \right\rangle_\qcutineq &\mbox{ for } n=1, \\[5pt]
 q_n(q^2_{\textrm{cut}})\equiv \left\langle (  q^2 - \left\langle q^2 \right\rangle )^n \right\rangle_\qcutineq & \mbox{ for } n>1,
\end{align}
which are related to the moments in~\eqref{eqn:q2momdef} via the binomial formula
\begin{equation}
  \left\langle (  q^2 - a )^n \right\rangle = 
  \sum_{i=0}^n 
  \binom{n}{i}
  \left\langle ( q^2)^i \right\rangle (-a)^{n-i}.
\end{equation}

\begin{figure}[htb]
  \centering
  \includegraphics[width=0.45\textwidth]{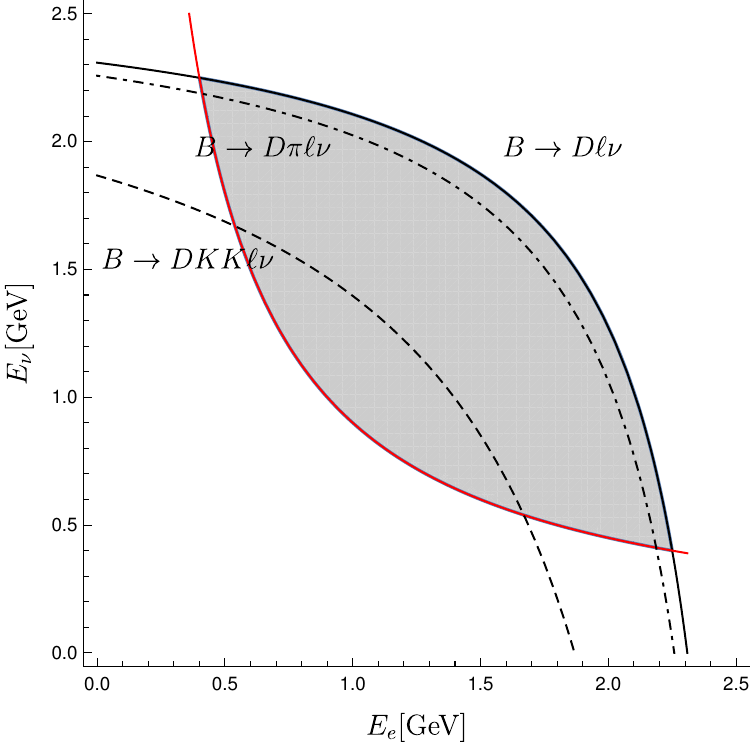} \quad
  \includegraphics[width=0.45\textwidth]{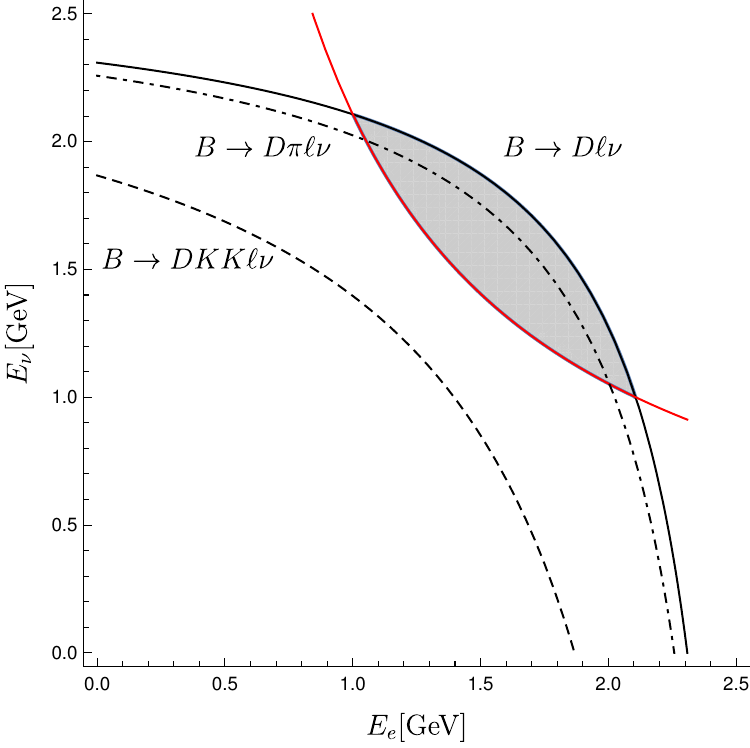}
  \caption{Gray areas are the kinematically allowed regions in the $E_\ell$-$E_{\nu}$ plane where $B \to X_c \ell \nu$ events can acquire a leptonic invariant mass larger than $\qcut = 3.60$ GeV$^2$ (left) and $q^2_\text{cut} = 8.43$ GeV$^2$ (right). Decay modes which require a hadronic invariant mass larger than $m_D$ populate the plots in areas closer to the origin, for instance, $B \to D \pi \ell \nu$ and $B \to D KK \ell \nu$ events only appear below the dot-dashed and dashed lines, respectively. 
  }
  \label{fig:EeEvPS}
\end{figure}
To further study the effect of the $q^2$ cut, we show in Fig.~\ref{fig:EeEvPS} the allowed phase space for a $B \to X_c \ell \nu$ decay in the $E_\ell$-$E_\nu$ plane with $\qcut=3.6$ GeV$^2$ and $\qcut=8.4$ GeV$^2$. 
The phase space is limited from below by
\begin{equation}
  \frac{\qcut}{4 E_\ell} \le E_\nu , 
\end{equation}
and all events with $q^2>\qcut$ lie above this curve.
This illustrates that the cut on $q^2$ removes all the events with low lepton energy $E_\ell$, therefore a cut on $q^2$ can replace 
a cut on $E_\ell$.
In addition, 
\begin{equation}
  m_D^2 \le m_X^2 = m_B^2 -2 m_b (E_\ell+E_\nu)+ 2 E_\ell E_\nu (1-\cos \theta_{\ell\nu}) \le m_B^2 -2 m_b (E_\ell+E_\nu)+ 4 E_\ell E_\nu ,
\end{equation}
determines the upper limit of the phase space in Fig.~\ref{fig:EeEvPS}.
Increasing the value of $\qcut$ makes the inclusive $B \to X_c \ell \nu$ measurement less and less inclusive, as is illustrated by the dot-dashed and dashed lines, which show where the $B \to D\pi\ell\nu$ and $B\to DKK\ell \nu$ modes populate the plot, respectively. Hence the situation is similar to a cut in the lepton energy.

Rewriting the phase space equations, we find 
\begin{equation}
  E_\ell \ge \frac{m_b^2+\qcut-m_D^2 - \lambda^{1/2}(m_B^2,\qcut,m_D^2)}{4 m_B} .
  \label{eqn:emin}
\end{equation}
where $\lambda(m_B^2, \qcut, m_D^2)$ is the K\"{a}ll\'{e}n function.  
Therefore, if $\qcut$ is chosen to be larger than the critical value
\begin{equation}
  q^2_\text{cr} = 2 \ecut m_B - \frac{2 \ecut m_D^2}{m_B-2 \ecut},
\end{equation}
then the $q^2$-moments do not depend $\ecut$ since the lepton energy always respects $\ecut \le E_\ell $ via the 
constraint~\eqref{eqn:emin}. A typical $\ecut = 0.4$ GeV, would then correspond to $\qcut = 3.6 $ GeV$^2$.

\subsection{Extracting \boldmath $V_{cb}$ \unboldmath}
\label{sec:extraction}
We propose to extract $V_{cb}$ based on the reduced set of HQE parameters, using measurements of the total rate and the $q^2$ moments, including a cut on $q^2$.  This strategy is identical to the approach in Ref.~\cite{Gam16}, however with theoretical expressions depending on fewer parameters (even for a fit including terms up to $1/m_b^3$ only). 
Since there are only five additional parameters at order $1/m_b^4$, precise input from the $q^2$ spectrum would allow us to perform a fully data driven analysis, i.e.\ an extraction of the HQE parameters at $1/m_b^4$ entirely from data.    

To this end, the expressions for the $q^2$ moments can be inverted to extract the HQE parameters. Therefore, it is important that the $(q^2)^n$  moments depend in a different way on the HQE parameters for different values of $n$. To show explicitly that the first four moments are linearly independent, we give the centralized moments in Eq.~\eqref{eq:cenmom} for $q^2_{\rm cut}=0$ GeV:

\vspace{0.0cm}
{\footnotesize
\begin{align}
  q_1{}& =\frac{m_b^2}{\mu_3}\left(0.23 \mu_3 -0.59 \frac{\mu_G^2}{m_b^2}-1.4\frac{(\mu_G^2)^2}{m_b^4 \mu_3}-6.0\frac{\tilde\rho_D^3}{m_b^3} +17\frac{r_E^4}{m_b^4}-6.2\frac{r_G^4}{m_b^4}  -1.9\frac{s_E^4}{m_b^4}+0.26\frac{s_B^4}{m_b^4} -0.072\frac{s_{qB}^4}{m_b^4}\right) \, ,\nonumber \\[5pt]
  q_2{}& = \frac{m_b^4}{\mu_3} \left(0.024 \mu_3 -0.13 \frac{\mu_G^2}{m_b^2}-0.66\frac{(\mu_G^2)^2}{m_b^4 \mu_3}-1.9\frac{\tilde\rho_D^3}{m_b^3} +8.7\frac{r_E^4}{m_b^4}-2.4\frac{r_G^4}{m_b^4} -0.80\frac{s_E^4}{m_b^4}+0.31\frac{s_B^4}{m_b^4} -0.098\frac{s_{qB}^4}{m_b^4}\right) \, ,\nonumber \\[5pt]
q_3{}& = \frac{m_b^6}{\mu_3} \left(1.4 \cdot 10^{-3} \;\mu_3 -0.016 \frac{\mu_G^2}{m_b^2}-0.27\frac{(\mu_G^2)^2}{m_b^4 \mu_3}-0.42\frac{\tilde\rho_D^3}{m_b^3} +3.4\frac{r_E^4}{m_b^4}-0.69\frac{r_G^4}{m_b^4} \right.\nonumber \\
{}& \left. \quad\quad\quad\quad\quad -0.25\frac{s_E^4}{m_b^4}+0.14\frac{s_B^4}{m_b^4} -0.21\frac{s_{qB}^4}{m_b^4}\right) \, ,\nonumber \\[5pt]
q_4{}&  = \frac{m_b^8}{\mu_3} \left(1.2\cdot 10^{-3} \mu_3 -0.014 \frac{\mu_G^2}{m_b^2}-0.12\frac{(\mu_G^2)^2}{m_b^4 \mu_3}-0.28\frac{\tilde\rho_D^3}{m_b^3} +2.1\frac{r_E^4}{m_b^4}-0.42\frac{r_G^4}{m_b^4} \right.\nonumber \\[5pt]
{}& \left. \quad\quad\quad\quad\quad -0.15\frac{s_E^4}{m_b^4}+0.077\frac{s_B^4}{m_b^4} -0.024\frac{s_{qB}^4}{m_b^4}\right)\, .
\label{eqn:q2centnum}
\end{align}}
These expressions are obtained from Eq.~\eqref{eqn:q2momdef} by re-expanding the ratio in $1/m_b$ up to $1/m_b^4$. We observe that the higher moments are more sensitive to the higher order terms, as is also the case for the energy and hadronic mass moments.

\begin{figure}[htp]
\centering
\subfloat[]{\includegraphics[width=0.65\textwidth,valign=t]{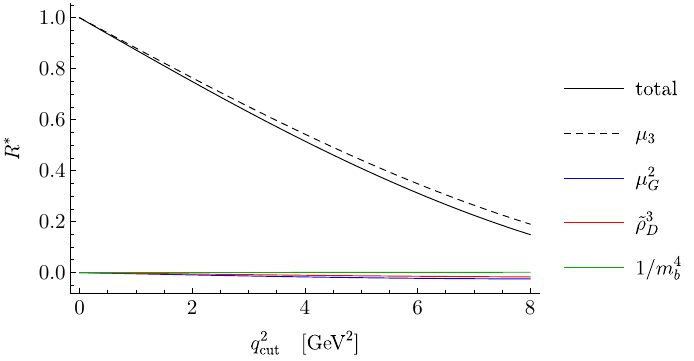}}  
\hfill\null
 \\
\subfloat[]{\includegraphics[width=0.5\textwidth]{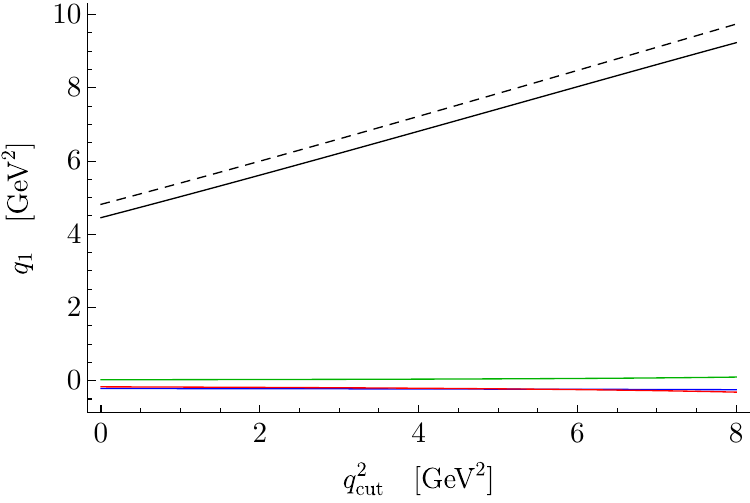}}
\subfloat[]{
\includegraphics[width=0.5\textwidth]{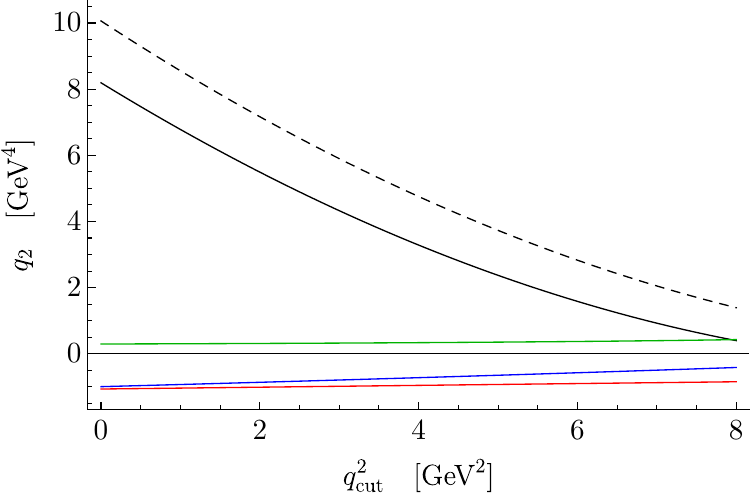}} \\
\subfloat[]{\includegraphics[width=0.5\textwidth]{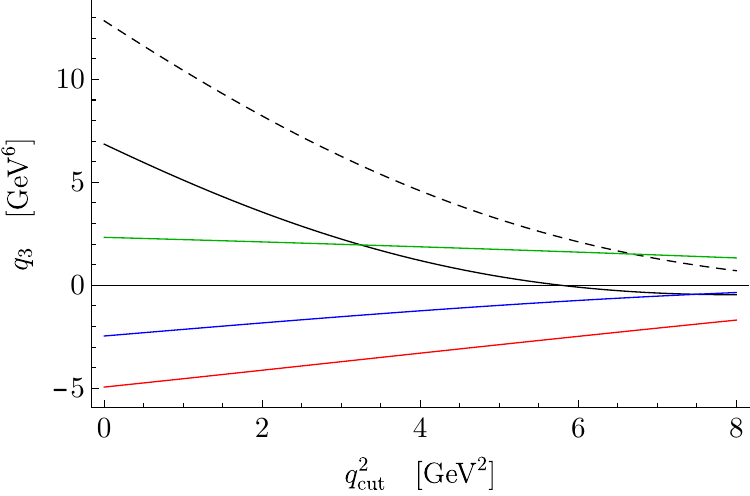}}
\subfloat[]{
\includegraphics[width=0.5\textwidth]{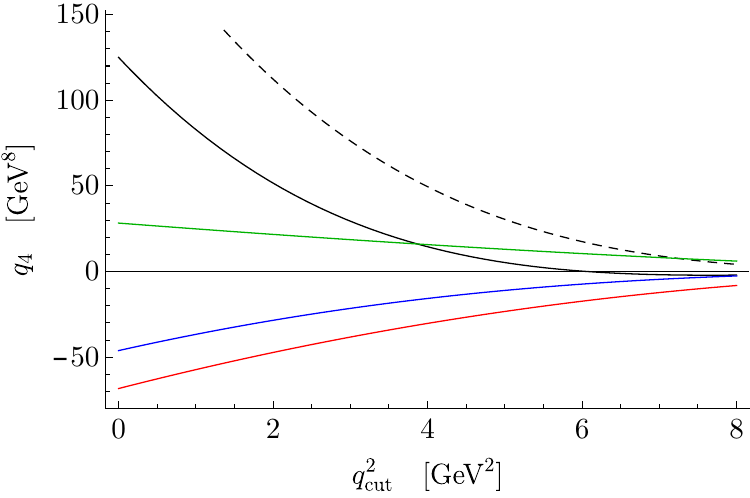}}
\caption{Dependence of the ratio $R^*$ and the first four centralized moments on the different orders in the $1/m_b$ expansion. Black dashed, blue and green lines represent the contributions from $\mu_3$, $\mu_G^2$ and $\tilde \rho_D^3$, respectively. The sum of the $1/m_b^4$ terms are depicted in green. Black solid lines are the final predictions for $R^*$ and the moments including terms up to order $1/m_b^4$. } 
\label{fig:centralmoments}
\end{figure}

Finally, as discussed, $V_{cb}$ could also be determined by making use of the $q^2_\mathrm{cut}$ dependence of the moments. In Fig.~\ref{fig:centralmoments}, we show the ratio $R^*$ and the centralized moments $q_{1,2,3,4}$ as a function of $q^2_{\textrm{cut}}$. We present their prediction up to $1/m_b^4$ together with the different contributions given by $\mu_3$, $\mu_G^2$, $\tilde \rho_D^3$ and the sum of the five operators of order $1/m_b^4$.
Also in this case, they are evaluated by re-expanding~\eqref{eqn:q2momdef} and~\eqref{eqn:Rstar} up to $1/m_b^4$. The size of the various $1/m_b$ terms is then obtained by selecting only the relative term in the expansion, multiplied by the residual $m_b^{2n}/\mu_3 \approx m_b^{2n}$, as in Eq.~\eqref{eqn:q2centnum}. 
Note that after the re-expansion there are terms proportional to $(\mu_G^2)^2/m_b^4$; they are much smaller than the genuine $1/m_b^4$ contribution and therefore not presented in Fig.~\ref{fig:centralmoments}. 
For the numerical values of the HQE parameters, we use those obtained in \cite{Gam16} as a benchmark scenario. We list the conversion of our basis to the one of Refs.~\cite{Gam16, Man10} in Appendix~\ref{sec:ap}. 

The expressions in Eq.~\eqref{eqn:q2centnum} and the plots presented in Fig.~\ref{fig:centralmoments}, show a good behaviour of the OPE expansion for the ratio $R^*$ and the first moment, which are dominated by the leading contribution proportional to $\mu_3$. 
Centralized moments of higher order on the contrary have a strong dependence on the higher order terms, since the subtraction $q^2-\langle q^2 \rangle$ removes the bulk of the contribution from $\mu_3$. For the third and fourth moments, higher order terms form a non-negligible fraction of total contribution.
As for the individual contributions of the $1/m_b^4$ terms to the centralized moments, we observe that the largest contribution of the $1/m_b^4$ comes from the parameter $r^4_E$, followed in size by $s^4_E,s^4_B$ and $s^4_{qB}$. On the contrary, $r^4_G$ gives a much smaller contribution to the moments. Due to the numerical values adopted for these parameters in Eq.~\eqref{eq:num4}, the suppression of the spin-dependent $s^4$ terms seen in Eq.~\eqref{eqn:q2centnum} is lifted and therefore the centralized moments seem sensitive also to the spin-dependent parameters.   
We emphasize that especially for values of $\qcut > 3$, which as discussed would correspond to a lepton energy cut of $E_\ell = 0.4$ GeV, their contribution becomes more pronounced, showing the feasibility of the proposed strategy in employing a $q^2$ cut. 

\section{Conclusion}
The extraction of $V_{cb}$ from inclusive decays based on the HQE provides a determination with a relative uncertainty of about 2\%; in combination with the exclusive measurement obtained from $B \to D^{(*)}  \ell \bar{\nu}$ and precise lattice data of the relevant form factors, $V_{cb}$ will be among the best known CKM matrix elements.  
 
Among the main theoretical challenges of the inclusive determination there is the proliferation of HQE parameters as soon as one includes higher orders in $1/m_b$, which complicates their extraction from data. 
As we have discussed in this paper, one may make use of reparametrization invariance, which leads to relations between different order in the HQE, to reduce the set of parameters for specific observables  like total rates, the leptonic invariant mass spectrum and its moments.
 
In the context of the inclusive $V_{cb}$ determination this may open the road to perform a purely data-driven analysis including also higher order terms. We discussed a new method based on the leptonic invariant mass spectrum and its moments. The proposed strategy is similar to the one pursued in the $V_{cb}$ fits relying on the charged lepton energy and the hadronic invariant mass moments. 
 
From the experimental side, the proposed strategy requires the reconstruction of the neutrino momentum which is possible only at $e^+ e^-$ colliders.  $B$-tagging algorithms are  well established methods at $B$ factories to identify $B \bar{B}$ events. They provide kinematical constraints that allows for a precise momentum reconstruction of the $B$ meson which decays semileptonically, and therefore the direction of the undetected neutrino can be constructed using energy conservation. Already with the existing $\Upsilon(4S)$ data from Belle~(711~fb$^{-1}$) and BABAR~(433~fb$^{-1}$) it should be possible the test our alternative method by measuring the $q^2$ moments and extracting $V_{cb}$.   
Finally, it would be interesting to perform a dedicated analysis of the new method at Belle-II in order to fully exploit our new method and to determine $V_{cb}$ in a fully data-driven method including $1/m_b^4$ corrections. 

\section*{Acknowledgements} 
We thank Phillip Urquijo and Florian Bernlochner for discussions related to this subject. 
This work was supported by DFG through the Research Unit FOR 1873 ''Quark Flavour Physics and Effective Field Theories''. 

\appendix
\section{Conversion Between Different Conventions} 
\label{sec:ap}
In the following, we discuss the conversion between the HQE parameter definitions in this work and those in~\cite{Man10, Gam16}. These $V_{cb}$ determinations are done with operators defined in a basis in which the covariant derivative is split into a spatial and a time derivative via 
\begin{equation}\label{eq:perpbasis}
i D_\mu  = v_\mu ivD + D_\mu^\perp \ .
\end{equation}
As we stressed before \cite{MannelVos}, this basis is less useful when considering RPI quantities. Changing between these bases  involves the absorption of higher-order terms. In particular, we find:
\begin{align}
(\mu^2_\pi)^\perp {}& = 2 m_b^2(1 - \mu_3) + \mu_G^2 - \frac{r^4_G}{8 m_b^2} - \frac{s^4_B}{
 4 m_b^2} + \frac{\delta^4_{G1}}{4 m_b^2} + \frac{\delta^4_{G2}}{4 m_b^2} \\
(\mu_G^2)^\perp {}& =  \mu_G^2 + \frac{\rho^3_D}{m_b} + \frac{\rho^3_{LS}}{m_b}  
\ , \label{eq:mugexp}\\ 
(\rho_{LS}^3)^\perp {}& =  \rho_{LS}^3 -\frac{r_E^4}{2m_b} - \frac{s_E^4}{2m_b} \ , \\
(\rho_D^3)^\perp {}&= \rho_D^3 \ .
\end{align}
Here, we have introduced four additional non-RPI parameters; 
\begin{subequations}\label{eq:MEsad}
\begin{eqnarray}
&&\frac{1}{2} \langle \bar{Q}_v  \left\{ iD_\alpha \, , \, \left[ ivD \, , \,  iD_\beta \, \right] \right\}(-i \sigma^{\alpha \beta}) Q_v 
\rangle = 2 m_B \rho_{LS}^3  \vphantom{\frac{1}{1} }\ ,  \\ 
&& \frac{1}{2}\langle \bar{Q}_v \left[ (iD_\mu) \, , \,   \left[ (iD)^2  \, , \, (i D^\mu) \right] \right]   Q_v \rangle 
= 2 m_B  \delta\rho_D^4  \vphantom{\frac{1}{1} }\ ,  \\ 
&&\frac{1}{2} \langle \bar{Q}_v  \left\{ iD_\alpha \, , \, \left[ (iD)^2 \, , \,  iD_\beta \, \right] \right\}(-i \sigma^{\alpha \beta}) Q_v 
\rangle = 2 m_B \delta\rho_{LS}^4  \vphantom{\frac{1}{1} }\ ,  \\ 
&& \langle \bar{Q}_v  (iD^2)^2 Q_v \rangle 
= 2 m_B \delta_{G1}^4  \vphantom{\frac{1}{1} }\ ,  \\ 
&& \langle \bar{Q}_v \left\{ (iD)^2 \, , \, \sigma\cdot G \right\}    Q_v \rangle 
= 2 m_B  \delta_{G2}^4  \vphantom{\frac{1}{1} }\ ,  
\label{eqn:G2}
\end{eqnarray} 
\end{subequations}
which do not occur in the total rate and the $q^2$ moments. 
Here 
\begin{equation}
\sigma \cdot G \equiv -i\sigma_{\mu\nu} (iD^\mu)(iD^\nu) \ , \;\;\; \gamma_\mu \gamma_\nu = g_{\mu\nu} + (-i\sigma_{\mu\nu}) \ .
\end{equation}
We note that, in Eq.~\eqref{eq:MEs}, we have defined $\tilde\rho_D^3$ including its RPI completion, where
\begin{equation}\label{eq:rhodexp}
\tilde\rho_D^3 = \rho_D^3 + \frac{1}{2m_b} \delta\rho_D^4 \ .
\end{equation} 
We emphasize that $\tilde\rho_D^3$ is sometimes defined without a commutator, which makes a difference at higher orders. 
Similar, the additional matrix elements in Eq.~\eqref{eq:MEsad} contain a completion of the $\rho_{LS}^3$ via
\begin{equation}
\tilde\rho_{LS}^3 \equiv \rho_{LS}^3 + \frac{1}{2m_b} \delta\rho_{LS}^4 \ .
\end{equation}

The 4th order parameters first introduced in Ref.~\cite{Man10} and determined in \cite{Gam16} are related to our parameters via
\begin{equation}
\begin{array}{llll}
  m_1 =& \frac{1}{3} \left( r_E^4 +  \frac{1}{2} r_G^4 + 2 \delta\rho_D^4 + \delta_{G1}^4 \right), & m_6  =& -s^4_B + s^4_E,\\
m_2 =& -r_E^4, &m_7 =& 2 \delta\rho_{LS}^4 + 2 s_E^4 + \frac{1}{2} s_{qB}^4, \\
m_3=& -2 r_E^4 + r_G^4, & m_8  =& 4 \delta_{G2}^4, \\
m_4 =& 2 r_E^4 - 2r_G^4 -2 \delta\rho_D^4, &m_9 = &-2 s_B^4 + 2 s_E^4 + \frac{1}{2} s_{qB}^4 \  . \\
m_5 =&  -s^4_{E}, & \\
\end{array}
\end{equation}

For our numerical analysis, we used these transformations to estimate our parameters from the determinations of the $m_i$ in \cite{Gam16}. We find 
\begin{equation}\label{eq:num4}
r_E^4 =0.019, \; r_G^4 = -0.006, \; s^4_E= -0.072, \; s^4_B= -0.13, \; s^4_{qB} = -0.80. 
\end{equation}
For the terms up to $1/m_b^3$, we use
\begin{equation}
\mu_3 = 1 + \frac{\mu_G^2 - \mu_\pi^2}{2 m_b^2} = 0.998 \, 
\end{equation}
and 
\begin{equation}
\mu_G^2 = 0.362%
, \;\;\; \text{and}\;\;\; \tilde\rho_D^3 = 0.127.%
\end{equation}
which we obtained by inserting the values found in \cite{Gam16} in Eq.~\eqref{eq:mugexp} and \eqref{eq:rhodexp}, respectively.
In addition, we used \cite{Gam16}
\begin{equation}
m_b^{\textrm{kin}} = 4.546, \; \;\textrm{and}\;\; \bar{m}_c(3 \textrm{GeV}) = 0.987 \ .
\end{equation}

For completeness, we also give the total rate in terms of our matrix elements; 
\begin{align}
\frac{\Gamma}{\Gamma_0} &= 
\mu_3 \; z(\rho) 
-2 \, \frac{\mu_G^2}{m_b^2}  (\rho -1)^4 + d(\rho) \; \frac{\tilde\rho_D^3}{m_b^3} + \frac{2}{3} (-1+\rho)^3 (1+5\rho) \frac{s^4_B}{m_b^4}  \nonumber\\[5pt]
{}& -\frac{8}{9} \frac{r_E^4}{m_b^4} \Big( 2 + 9 \rho^2 - 20 \rho^3 + 9 \rho^4 + 6 \log \rho \Big)\nonumber \\[5pt]
{}&+ \frac{4}{9} \frac{r_G^4}{m_b^4} \Big( 16-21 \rho +9 \rho^2 - 7\rho^3 + 3\rho^4 +12 \log \rho \Big)\nonumber\\[5pt]
{}&+\frac{1}{9} \frac{s_E^4}{m_b^4}  \Big(50 - 72 \rho + 40 \rho^3 - 18 \rho^4 + 24 \log \rho \Big)\nonumber \\[5pt]
{}&+\frac{1}{36} \frac{s^4_{qB}}{m_b^4}  \Big(-25 + 48 \rho - 36 \rho^2 + 16 \rho^3 - 3 \rho^4 - 12 \log \rho \Big)+ \mathcal{O}(1/m_b^5) \ , 
\end{align}

where
\begin{align}
z(\rho)\equiv {}&  1-8\rho + 8 \rho^3 - \rho^4 -12\rho^2 \log\rho \, ,\\
d(\rho)\equiv {}& \frac{2}{3} \Big(17 - 16 \rho - 12  \rho^2 + 16 \rho^3 - 5 \rho^4 +  12 \log\rho \Big) \, .
 \end{align}

\end{document}